\documentclass[preprint,aps,prd,showpacs,superscriptaddress,nofootinbib,tightenlines]{revtex4}

\usepackage{amsfonts}
\usepackage{multirow}
\usepackage{mathrsfs}
\usepackage{graphicx}
\usepackage{amsmath}
\usepackage{amssymb}
\usepackage{bm}
\usepackage{bbm}
\usepackage{color}
\usepackage{array}

\usepackage{diagbox}
\usepackage{ulem}
\usepackage{slashbox}

\def\pslash{p\!\!\!\slash}
\def\qslash{q\!\!\!\slash}

\def\OMIT#1{}

\newcommand{\nn}{\nonumber}

\newcommand{\beq}{\begin{equation}}
\newcommand{\eeq}{\end{equation}}
\newcommand{\bqa}{\begin{eqnarray}}
\newcommand{\eqa}{\end{eqnarray}}

\begin{document}

\title{\mbox{}\\[10pt]
Study on $\eta_{c2}(\eta_{b2})$ electromagnetic decay into double photons}
\author{Li Yang}
 \affiliation{School of Physical Science and Technology, Southwest University, Chongqing 400700, China\vspace{0.2cm}}

\author{Wen-Long Sang~\footnote{wlsang@swu.edu.cn}}
 \affiliation{School of Physical Science and Technology, Southwest University, Chongqing 400700, China\vspace{0.2cm}}

\author{Hong-Fei Zhang~\footnote{shckm2686@163.com}}
\affiliation{College of Big Data Statistics, Guizhou University of Finance and Economics, Guiyang, 550025, China\vspace{0.2cm}}

\author{Yu-Dong Zhang}
 \affiliation{School of Physical Science and Technology, Southwest University, Chongqing 400700, China\vspace{0.2cm}}

\author{Ming-Zhen Zhou~\footnote{zhoumz@swu.edu.cn}}
 \affiliation{School of Physical Science and Technology, Southwest University, Chongqing 400700, China\vspace{0.2cm}}
\begin{abstract}
Within the framework of nonrelativistic QCD (NRQCD) factorization formalism, we compute the helicity amplitude as well as the decay width of $\eta_{Q2}$ ($Q=c,b$) electromagnetic decay into two photons up to next-to-next-to-leading order (NNLO) in $\alpha_s$ expansion. For the first time, we verify the validity of NRQCD factorization for the D-wave quarkonium decay at NNLO. We find that the $\mathcal{O}(\alpha_s)$ and $\mathcal{O}(\alpha_s^2)$ corrections to the helicity amplitude are negative and moderate, nevertheless both corrections combine to
suppress the leading-order prediction for the decay width significantly.
By approximating the total decay width of $\eta_{Q2}$
 as the sum of those for the hadronic decay and the electric $E1$ transition, we obtain the branching ratios ${\rm Br}(\eta_{c2}\to 2\gamma)\approx 5\times10^{-6}$ and ${\rm Br}(\eta_{b2}\to 2\gamma)\approx 4\times10^{-7}$. To explore the potential measurement on $\eta_{Q2}$,
we further
evaluate the production cross section of $\eta_{Q2}$ at LHCb at the lowest order in $\alpha_s$ expansion. With the kinematic constraint on the longitudinal rapidity $4.5>y>2$ and transverse momentum $P_T>(2-4)m_Q$ for $\eta_{Q2}$, we find the cross section can reach $2-50$ nb for $\eta_{c2}$, and $1-22$ pb for $\eta_{b2}$. Considering the integrated luminosity $\mathcal{L}=10\, {\rm fb}^{-1}$ at $\sqrt{s}=7$ TeV and $\sqrt{s}=13$ TeV, we estimate that there are several hundreds events of $pp\to \eta_{c2}\to 2\gamma$. Since the background is relatively clean, it is promising to reconstruct $\eta_{c2}$ through its electromagnetic decay. On the contrary, due to small branching ratio and production cross section, it is quite challenging to detect $\eta_{b2}\to 2\gamma$ at LHCb.
\end{abstract}

\keywords{Quarkonium, NRQCD, decay}

\maketitle


\section{introduction}
Heavy quarkonium, as a multiscale system, is an ideal laboratory for testing the interplay
between perturbative and nonperturbative QCD. Its mass spectrum has been predicted by various potential models, and
most of the low-lying quarkonium states have been probed by the experiment. However there still exit some undiscovered states.
Among the missing states in charmonium family,
$\eta_{c2}(^1D_2)$ is the only spin-singlet low-lying D-wave state.
A full understanding on $\eta_{c2}$ in both theory and experiment can help to illuminate the interquark force and reveal the nature of the strong interaction.

The mass of $\eta_{c2}$ is predicted to range from $3.80$ to $3.88$ GeV~\cite{Eichten:2004uh,Barnes:2005pb,Li:2009zu,Godfrey:1985xj,Fulcher:1991dm,Zeng:1994vj,Ebert:2002pp}, which well lies between the $D\bar{D}$ and the $D^*\bar{D}$ thresholds.  Quite different from $\psi''$, the decay of $\eta_{c2}$ into $D\bar{D}$ is forbidden, which is accounted for by the parity conservation. Thus $\eta_{c2}$ is
a narrow resonance, and its main decay modes are considered to be hadronic decay and electric $E1$ transition, which have been well investigated in the references.

The electric $E1$ transition has been known for a long time~\cite{Eichten:2002qv,Ebert:2002pp,Barnes:2005pb} (see also the review in \cite{Brambilla:2004wf}).
The hadronic transition $\eta_{c2}\to \pi\pi\eta_c$ was evaluated in Ref.~\cite{Eichten:2002qv}.
The $\eta_{c2}$ hadronic decay was studied in Refs.~\cite{Novikov:1977dq,Fan:2009cj}.
Production through $B$ meson decay is an important channel to search for charmonia~\cite{Choi:2002na,Chilikin:2019wzy,Abe:2001pfa,Aubert:2002ht,Aubert:2001xs,Edwards:2000bb}.
Based on the  nonrelativisitc QCD (NRQCD) factorization formalism~\cite{Bodwin:1994jh}, the inclusive $\eta_{c2}$ production in B decay was evaluated and proposed to probe $\eta_{c2}$ through this channel~\cite{Ko:1997rn,Fan:2011aa}. The decay $B^-\to \eta_{c2}K^-$ has been explored by using the
rescattering mechanism in Ref.~\cite{Xu:2016kbn}.  In addition, the $\eta_{Q2}$ electromagnetic decay into double photons
was evaluated by using the instantaneous Bethe-Salpeter method in Ref.~\cite{Wang:2013nya}. Unfortunately, no significant signal has
been found till today~\cite{Chilikin:2020irx}.

The $C=+1$ charmonia bear a considerable branching ratio of electromagnetic decay into double photons, e.g.,
${\rm Br}(\eta_c\to 2\gamma)\approx 1.57\times 10^{-4}$, ${\rm Br}(\chi_{c0}\to 2\gamma)\approx 2.04\times 10^{-4}$, and ${\rm Br}(\chi_{c2}\to 2\gamma)\approx 2.85\times 10^{-4}$~\cite{Tanabashi:2018oca}, which have been well measured by the experiment.
Although the electric $E1$ transition $\eta_{c2}\to h_c\gamma$ comprises one of the main decay channel,
it is anticipated that the branching ratio for $\eta_{c2}$ electromagnetic decay is of the same magnitude as those in $\eta_{c}$ and $\chi_{c}$. Compared with the hadronic decay, the electromagnetic decay has the advantage of bearing a clean background.
Thus we propose to detect $\eta_{c2}$ through $\eta_{c2}\to 2\gamma$.
In this work, we will evaluate the partial width of $\eta_{Q2}\to 2\gamma$ (the heavy quark flavor $Q=c,b$) within framework of
the well established NRQCD factorization formalism.

NRQCD factorization formalism is widely employed to tackle heavy quarkonium decay and production. Within this framework,
the production cross section or decay width can be systematically disentangled the short-distance and long-distance effect, formalized by a double expansion in powers of heavy quark velocity $v_Q$ and strong coupling constant $\alpha_s$. The perturbative contributions with the scale larger than the heavy quark $m_Q$ is encoded into the short-distance coefficients (SDCs),
while the nonperturbative effects are depicted by the NRQCD long-distance matrix elements (LDMEs).

Recently, there is a remarkable progress in deducing the higher-order perturbative corrections for various quarkonium decay and production processes~\cite{Czarnecki:1997vz,Beneke:1997jm,Czarnecki:2001zc,Onishchenko:2003ui,Marquard:2014pea,Beneke:2014qea,Chen:2015csa,
Sang:2015uxg,Feng:2015uha,Chen:2017soz,Chen:2017pyi,Feng:2017hlu,Wang:2018lry,Feng:2019zmt,Kniehl:2019vwr,Yu:2020tri,Sang:2020fql}.
It has been found even though the $\mathcal{O}(\alpha_s)$ corrections to the charmonium electromagnetic decay are moderate,
the $\mathcal{O}(\alpha_s^2)$ corrections can be considerable. Therefore it is mandatory to include the $\mathcal{O}(\alpha_s^2)$ contributions in our evaluation. Concretely speaking, we will evaluate the decay width of $\eta_{Q2}\to 2\gamma$ up to NNLO in $\alpha_s$ expansion.

In addition, to provide aid for experimental search for $\eta_{Q2}$ through $\eta_{Q2}\to 2\gamma$,
we will explore the $\eta_{Q2}$ production at colliders, and estimate the corresponding number of events.
The $\eta_{c2}$ associated production with a photon at B factory has been obtained in Ref.~\cite{Li:2009ki}.
The $\eta_{c2}+J/\psi$ production at B factory can be found in Ref.~\cite{Braaten:2002fi}. Unfortunately, the cross sections of both channels are too small to probe $\eta_{c2}\to 2\gamma$.
Considering the significant luminosity as well as the sizeable cross sections for quarkonia production at LHC,  exemplified by $\sigma(pp\to \eta_c+X)\sim 0.5$ $\mu$b with the transverse momentum cut $p_T>4m_c$~\cite{Zhang:2014ybe}, we anticipate the cross section for $\eta_{c2}$ production is also considerable.
Actually, the cross section of $\eta_{c2}$ at LHC can be simply estimated by $\sigma(\eta_{c2})\sim \frac{|\mathcal{R}^{''}_{D}(0)|^2}{m_c^4|\mathcal{R}_{S}(0)|^2}\times\sigma(\eta_c)\sim v_c^4\times 0.5 \sim 5$ nb.
Thus, we expect there are amount number of events for $pp\to \eta_{c2}\to 2\gamma$, which renders $\eta_{c2}\to 2\gamma$ a promising channel to probe $\eta_{c2}$.

The remainder of this paper is organized as follows. In Sec.~\ref{sec-formulism}, we make Lorentz decomposition for the amplitude
of $\eta_{Q2}\to 2\gamma$, and present the decay width in term of the helicity amplitudes.
In Sec.~\ref{sec-nrqcd}, we outline the NRQCD factorization formalism for the helicity amplitude, and
briefly describe the theoretical framework to deduce the NRQCD SDC.
In Sec.~\ref{sec-SDC}, we first
introduce the technicalities encountered in performing loop calculation, and then present our final results for the SDC.
Sec.~\ref{sec-result-and-discussion} is devoted to the phenomenological analysis and discussion.
A theoretical prediction on the production cross section of $\eta_{Q2}$ at LHCb is also contained in this section. We present our
summary in Sec.~\ref{sec-summary}.

\section{Theoretical formula for the decay width\label{sec-formulism}}

The partial width of $\eta_{Q2}\to 2\gamma$ can be expressed in term of helicity amplitude
\bqa
\label{decay-rate}
\Gamma_{\gamma\gamma}(\eta_{Q2})=\frac{1}{2J+1}\frac{1}{2!}\frac{1}{8\pi}\bigg[2|\mathcal{A}_{1,1}|^2+2|\mathcal{A}_{1,-1}|^2\bigg],
\eqa
where $J=2$ denotes the spin of $\eta_{Q2}$, $\tfrac{1}{2!}$ accounts for the indistinguishability of the two identical photons, $\tfrac{1}{8\pi}$
corresponds to the phase space factor, and $\mathcal{A}_{\lambda_1,\lambda_2}$ signify the helicity amplitudes of $\eta_{Q2}\to \gamma(\lambda_1)\gamma(\lambda_2)$ with $\lambda_{1,2}=\pm 1$ being the helicity of the photons. By invoking the parity invariance, we only enumerate the independent helicity amplitudes~\footnote{Note that there is the relation $\mathcal{A}_{\lambda_1,\lambda_2}=(-1)^J\mathcal{A}_{\lambda_2,\lambda_1}$ for the two identical photons and the P-parity  further enforce restriction $\mathcal{A}_{\lambda_1,\lambda_2}=-\mathcal{A}_{-\lambda_1,-\lambda_2}$ for $\eta_{Q2}$~\cite{Haber:1994pe}, therefore we have $\mathcal{A}_{1,-1}=-\mathcal{A}_{1,-1}=0$. We will demonstrate the vanishment of $\mathcal{A}_{1,-1}$ through explicit evaluation in the following.}.

To extract the helicity amplitudes, we first decompose the amplitude $\mathcal{A}$ of $\eta_{Q2}\to 2\gamma$ by Lorentz invariance. By Bose symmetry, transversality for the polarization of photon, together with parity conservation,
the amplitude  can be generically expressed as
\bqa
\label{lorentz-decomposition}
\mathcal{A}&=&\frac{c_1}{m_Q^4}\epsilon^{\alpha\beta\rho\sigma}\epsilon^*_{1\alpha}\epsilon^*_{2\beta}k_{1\rho}p_{\sigma}
k_{1\mu}k_{1\nu}\epsilon_H^{\mu\nu}
+\frac{c_2}{m_Q^2}\epsilon^{\alpha\beta\mu\rho}\epsilon^*_{1\alpha}\epsilon^*_{2\beta}p_\rho\epsilon_{H\mu\nu}k_1^\nu\nn\\
&+&\frac{c_3}{m_Q^2}(\epsilon^{\alpha\mu\rho\sigma}\epsilon_{1\alpha}^*k_{1\rho}p_\sigma
\epsilon_{H\mu\nu}\epsilon_{2}^{*\nu}-\epsilon^{\alpha\mu\rho\sigma}\epsilon_{2\alpha}^*k_{1\rho}p_\sigma
\epsilon_{H\mu\nu}\epsilon_{1}^{*\nu}),
\eqa
where $p$ denotes half of the momentum of $\eta_{Q2}$, $k_1$ and $k_2$ signify the momenta of the two outgoing photons, $c_i$ ($i=1,2,3$) are Lorentz invariant and refer to form factors of the corresponding Lorentz structure, $\epsilon_1$ and $\epsilon_2$ represent the polarization vectors of the photons, and $\epsilon_H$ denotes the polarization tensor of $\eta_{Q2}$.

It is straightforward to deduce the helicity amplitudes $\mathcal{A}_{\lambda_1,\lambda_2}$ from Eq.~(\ref{lorentz-decomposition}).
To carry out the calculation, it is convenient to
construct the explicit expressions of the polarization tensor $\epsilon_H$ and polarization vectors $\epsilon_1$ and $\epsilon_2$~\cite{Sang:2009jc}.
We define the polarization vectors
\bqa\label{eq-polarization-vectors}
\epsilon_{+}^\mu=\frac{1}{\sqrt{2}}(0,-1,-i,0),\;\;\epsilon_{-}^\mu=\frac{1}{\sqrt{2}}(0,+1,-i,0),\;\;
\epsilon_{0}^\mu=(0,0,0,1).
\eqa
The five polarization tensors of $\eta_{Q2}$ are readily expressed as
\begin{subequations}\label{eq-polarization-tensor}
\bqa
\epsilon_{H\pm2}^{\mu\nu}&=&\epsilon^\mu_{\pm}\epsilon^\nu_{\pm},\\
\epsilon_{H\pm1}^{\mu\nu}&=&\frac{1}{\sqrt{2}}(\epsilon^\mu_{\pm}\epsilon^\nu_0+\epsilon^\mu_{0}\epsilon^\nu_{\pm}),\\
\epsilon_{H\pm0}^{\mu\nu}&=&\frac{1}{\sqrt{6}}(\epsilon^\mu_{+}\epsilon^\nu_{-}+2\epsilon^\mu_{0}\epsilon^\nu_{0}
+\epsilon^\mu_{-}\epsilon^\nu_{+}).
\eqa
\end{subequations}
If assuming the photon with momentum $k_1$ is outgoing in the positive $z$ direction,
its polarization vector $\epsilon_1$ equals to $\epsilon_{+}$ for helicity $+1$, and $\epsilon_{-}$ for helicity $-1$,
while the helicity polarization vector of the backward photon is just reversed.

Substituting the explicit expressions Eq.~(\ref{eq-polarization-vectors}) and Eq.~(\ref{eq-polarization-tensor}) into Eq.~(\ref{lorentz-decomposition}),
we obtain the helicity amplitudes in term of the three form factors
\begin{subequations}\label{helicity-amp-formula}
\bqa
\mathcal{A}_{1,1}&=&-\frac{2i}{\sqrt{6}}(c_1-c_2+c_3),\\
\mathcal{A}_{1,-1}&=&0,
\eqa
\end{subequations}
where $\mathcal{A}_{1,-1}$ explicitly vanishes.

\section{NRQCD factorization formalism for the helicity amplitude \label{sec-nrqcd}}

Owing to the strong interaction inner the hadron $\eta_{Q2}$, the helicity amplitude $\mathcal{A}_{1,1}$ is nonperturbative. Fortunately, we can employ the NRQCD factorization formalism to factorize the helicity amplitude into~\cite{Bodwin:1994jh}
\bqa
\label{nrqcd-factorization}
\mathcal{A}_{1,1}=\mathcal{C}_{1,1}(\mu_F)\frac{\langle0|\chi^{\dagger}{\mathcal K}_{^1D_{2}} \psi(\mu_F)|\eta_{Q2}\rangle}{m_Q^{5/2}}(1+\mathcal{O}(v^2)),
\eqa
where $\mathcal{C}_{1,1}$ represents the perturbative SDC which depicts a heavy quark pair annihilation into double photons, $\mu_F$ signifies the factorization scale, and
\bqa
{\mathcal K}_{^1D_{2}}=(-{i\over 2})^2(\overleftrightarrow{D}^i
\overleftrightarrow{D}^j-\tfrac{1}{3}\overleftrightarrow{\bm D}^2\delta^{ij})\epsilon_H^{ij}
\eqa
with $\epsilon_H$ being the polarization tensor of $\eta_{Q2}$.
LDME $\langle0|\chi^{\dagger}{\mathcal K}_{^1D_{2}} \psi(\mu_F)|\eta_{c2}\rangle$ deciphering the nonperturbative effect in the hadron is process-independent, and can be related to the second derivative of the radial wave function at the origin through
\bqa
\label{nrqcd-potential}
\langle0|\chi^{\dagger}{\mathcal K}_{^1D_{2}} \psi|\eta_{Q2}\rangle
\approx\sqrt{\frac{5N_c}{8\pi}}\mathcal{R}^{''}_{D}(0).
\eqa

To get the helicity amplitude, we must determine the SDC.
Since the SDC is irrelevant to the nonperturbative hadronization effect, it can be computed through the
standard matching technique. Concretely, we can replace the physical meson $\eta_{Q2}$ with a heavy quark pair $Q\bar{Q}$,
carrying the same quantum number as $^1D_2$.
The factorization formalism is also valid to the free quark state $Q\bar{Q}(^1D_2)$, therefore
after the replacement, Eq.~(\ref{nrqcd-factorization}) becomes to
\bqa
\label{nrqcd-factorization-pert}
\mathcal{A}_{1,1}(^1D_2)=\mathcal{C}_{1,1}\frac{\langle0|\chi^{\dagger}{\mathcal K}_{^1D_{2}} \psi|Q\bar{Q}(^1D_2)\rangle}{m_Q^{5/2}},
\eqa
where we have suppressed the factorization scale and the high-order relativistic corrections. The SDC in Eq.~(\ref{nrqcd-factorization-pert})
is exactly the same as in Eq.~(\ref{nrqcd-factorization}).
Since the amplitude $\mathcal{A}_{1,1}$ and matrix element are now perturbative, both sides of Eq.~(\ref{nrqcd-factorization-pert}) are calculable.
In principle, one can solve the SDC $\mathcal{C}_{1,1}$ in any prescribed $\alpha_s$ order.

In the following, we briefly describe the procedure to evaluate the perturbative helicity amplitude $\mathcal{A}_{1,1}(^1D_2)$.
We assign the momenta of the $Q$ and $\bar{Q}$ quarks to be
\bqa
\label{def-momenta}
p_1&=&p+q,\nn\\
p_2&=&p-q,
\eqa
where $p$ and $q$ represent half of the total momentum and relative momentum of $Q\bar{Q}$ pair, respectively. The
on-shell condition enforces that
\bqa
\label{on-shell-relation}
p^2=p_1^2&=&p_2^2=m_Q^2,\nn\\
p\cdot q=0.
\eqa

In our calculation, we first evaluate the amplitude $\mathcal{A}$ of $Q\bar{Q}\to 2\gamma$, then
employ the covariant spin-projector to extract the spin-singlet pattern of $Q\bar{Q}$.
To be consistent with the decay width formula (\ref{decay-rate}),
we utilize the nonrelativistically normalized spin-singlet/color-singlet projector~\cite{Bodwin:2013zu}, which reads
\bqa
\label{spin-projector}
\Pi_0=\frac{(\pslash+\qslash+m_Q)(\pslash+m_Q)\gamma_5(\pslash-\qslash-m_Q)}{8\sqrt{2}m_Q^3}\otimes {{\bf 1}_c \over \sqrt{N_c}}.
\eqa
The $L=2$ orbital partial wave can be projected out by differentiating the color-singlet/spin-singlet
quark amplitude with respect to the relative momentum $q$, followed by setting $q$ to zero
\bqa
\label{d-wave-amp}
\mathcal{A}(^1D_2)=\epsilon_{H\mu\nu}\frac{|\bf q|^2}{2!}\frac{\rm \partial^2}{{\rm \partial}q_\mu {\rm \partial}q_\nu}{\rm Tr}[\Pi^0\mathcal{A}]|_{q=0}.
\eqa

Now, we have collected all the necessary ingredients to calculate the amplitude of $Q\bar{Q}(^1D_2)\to 2\gamma$. Subsequently,
we can pick up the Lorentz invariant form factor $c_i$ through Eq.~(\ref{lorentz-decomposition}), and obtain the perturbative
helicity amplitude with the aid of Eq.~(\ref{helicity-amp-formula}). Meanwhile, the perturbative NRQCD matrix element
$\langle0|\chi^{\dagger}{\mathcal K}_{^1D_{2}}\psi|Q\bar{Q}(^1D_2)\rangle$ can also be carried out at desired $\alpha_s$ order.
At lowest order in $\alpha_s$, we have
\bqa
\label{ldme-pert}
\langle0|\chi^{\dagger}{\mathcal K}_{^1D_{2}}\psi|Q\bar{Q}(^1D_2)\rangle=\sqrt{2N_c}|\bf q|^2.
\eqa
Finally, it is straightforward to determine the SDC $\mathcal{C}_{1,1}$ at prescribed $\alpha_s$ order through Eq.~(\ref{nrqcd-factorization-pert}).

For the future convenience, we re-express the partial width of $\eta_{Q2}\to 2\gamma$ in term of SDC
\bqa
\label{decay-rate-1}
\Gamma_{\gamma\gamma}(\eta_{Q2})=
\frac{1}{5}\frac{1}{8\pi}|\mathcal{C}_{1,1}|^2\frac{|\langle0|\chi^{\dagger}{\mathcal K}_{^1D_{2}} \psi|\eta_{c2}\rangle|^2}{m_Q^{5}}.
\eqa

\section{SDC up to NNLO\label{sec-SDC}}
In this section, we first describe the computational technicalities utilized to evaluate the perturbative amplitude in detail,
then present our main results for the SDC $\mathcal{C}_{1,1}$.

\begin{figure}[htbp]
 	\centering
 \includegraphics[width=0.8\textwidth]{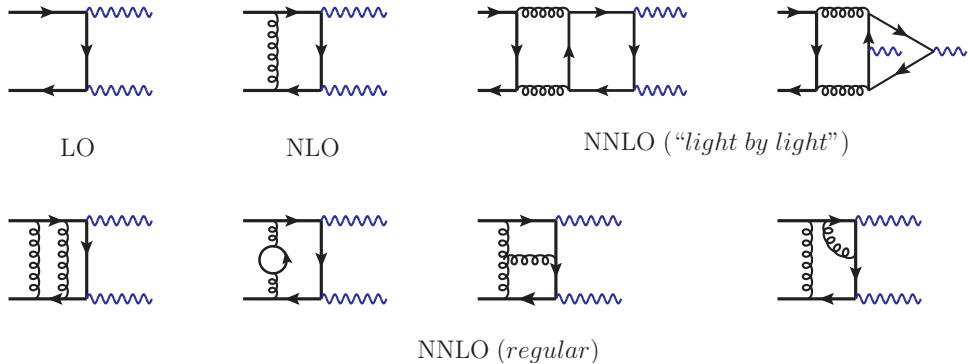}
 \caption{The representative Feynman diagrams for $Q\bar{Q}(^1D_2)\to 2\gamma$ through order $\alpha_s^2$.
 \label{feynman-fig-etac2}}
 \end{figure}

We employ {\tt FeynArts}~\cite{Hahn:2000kx} to generate the Feynman diagrams and the corresponding amplitude for
$Q\bar{Q}\to 2\gamma$. The representative Feynman diagrams are illustrated in Fig.~\ref{feynman-fig-etac2}.
We employ the spin-singlet/color-singlet projector (\ref{spin-projector}) and apply the recipe as specified in (\ref{d-wave-amp})
to project out the intended amplitude for $Q\bar{Q}(^1D_2)\to 2\gamma$ up to two-loop level. Subsequently,
the packages {\tt FeynCalc}~\cite{Mertig:1990an,Shtabovenko:2016sxi} and {\tt FormLink}~\cite{Feng:2012tk,Kuipers:2012rf} are
employed to perform the Dirac trace and Lorentz contraction.

For the NLO and NNLO corrections, we carry out the derivative of the amplitude with regard to the relative momentum $q$
prior to perform loop integration, which amounts to directly extract the contribution from the hard region.
We employ the package {\tt Apart}~\cite{Feng:2012iq} and {\tt FIRE}~\cite{Smirnov:2014hma} to conduct partial fraction and the
corresponding integration-by-part (IBP) reduction. Finally, we have 3 one-loop master integrals (MIs) and 80 two loop MIs.
There exist some complex-valued two-loop integrals originating from the light-by-light (lbl) diagrams, which are relatively hard to carry out numerically.
It deserves mentioning that although the MIs encountered here are almost the same sets as in the processes $\eta_c\to 2\gamma$ and $\chi_{c}\to 2\gamma$, the computation complexity is much more involved. Since we have taken the second derivative of the amplitude with
 respect to the relative momentum $q$, some coefficients of the MIs are relatively larger as well as more divergent in $1/\epsilon$ expansion compared with those in $\eta_c$ and $\chi_{c}$ decay. Thus, to reach the desired precision,
 we must perform numerical integration over the MIs to higher accuracy.

For the real-valued MIs, we directly use {\tt CubPack}/{\tt HCubature}~\cite{CubPack,HCubature} to carry out the integration.
In contrast to the application of SD to the Euclidean region, the singularities encountered in the physical region lie
inside, rather than sit on, the integration boundary, which render the integration hard to be numerically evaluated.
To overcome this difficulty, we conduct integration contour deformation via the variable transformation prior to decomposing the sectors~\cite{Borowka:2012qfa},
and determine the integration contour through optimizing a set of contour parameters~\cite{Feng:2019zmt}.
For more technical details, we refer the readers to Refs.~\cite{Feng:2019zmt,Sang:2020fql}.

The lbl Feynman diagrams are both gauge invariant and free of any UV and IR divergences in sum.
On the contrary, there exist UV divergence at one loop, and both UV and IR divergences at two loop.
The UV divergence originates from the integration over the loop momentum, which can be eliminated through the standard renormalization  procedure. We implement the on-shell renormalization for the heavy quark wave function and mass up to $\mathcal{O}(\alpha_s^2)$~\cite{Broadhurst:1991fy,Melnikov:2000zc,Baernreuther:2013caa}, and $\overline{\rm MS}$ renormalization for the strong coupling constant. Thus, the ultimate amplitude is completely UV finite.
There remains a piece of unremoved IR divergence, nevertheless this IR pole can be factored into the NRQCD LDME,
 so that the NRQCD SDC becomes IR finite. As a consequence, both of the LDME and the corresponding
two-loop SDC $\mathcal{C}_{1,1}$ bear $\ln\mu_F$ dependence, nevertheless their product must be independent of factorization scale.

After some hard work, we finally obtain the SDC $\mathcal{C}_{1,1}$ expanded in power of
the strong coupling constant $\alpha_s$
\bqa
\label{helicity-amp-explicit}
\mathcal{C}_{1,1}&=&\frac{4\sqrt{6}\pi}{3\sqrt{m_Q}}\alpha e_Q^2\bigg\{1+C_F\frac{\alpha_s}{\pi}\Delta^{(1)}
+\frac{\alpha_s^2}{\pi^2}\bigg[C_F\frac{\beta_0}{4}\Delta^{(1)}\ln\frac{\mu_R^2}{m_Q^2}+\Delta^{(2)}\bigg]\bigg\},
\eqa
where $\alpha$ denotes the electromagnetic coupling constant, $e_Q$ signifies the electric charge of the heavy quark,
$\beta_0=\frac{11}{3}C_A-\frac{2}{3}(n_L+n_H)$ corresponds to the one loop coefficient of the QCD $\beta$ function,
where $n_H=1$, and $n_L$ signifies the number of the active quark flavor ($n_L=3$ for $\eta_{c2}$, and $n_L=4$ for $\eta_{b2}$).
The exact occurrence of the $\ln\mu_R^2$ is demanded by the renormalization group invariance.

$\Delta^{(1)}$ and $\Delta^{(2)}$ correspond to the $\mathcal{O}(\alpha_s)$ and $\mathcal{O}(\alpha_s^2)$ corrections to the
SDC. The expression of $\Delta^{(1)}$ is analytically obtained
\bqa
\Delta^{(1)}=\frac{3}{8}\pi^2-6\ln2-1.
\eqa
$\Delta^{(2)}$ can be expressed as
\bqa
\label{NNLO-separation}
\Delta^{(2)}=-\frac{\pi^2}{10} C_F(C_A+2C_F)\ln\frac{\mu_F}{m_Q}+\Delta^{(2)}_{\rm reg}+\Delta^{(2)}_{\rm lbl}.
\eqa
The coefficient of the factorization scale dependence $\ln\mu_F$ term corresponds to the anomalous dimension of
the NRQCD operator in Eq.~(\ref{nrqcd-factorization}), which is consistent with the result in Ref.~\cite{Hoang:2006ty}~\footnote{In Ref.~\cite{Hoang:2006ty},
the authors computed the anomalous dimensions of spin-single and spin-triplet currents for heavy quark pair with arbitrary orbital angular
momentum. The anomalous dimension of $^1D_2$ can be obtained by utilizing Eq.~(40) in Ref.~\cite{Hoang:2006ty}
with color-singlet Wilson coefficients of the potentials given in Refs.~\cite{Manohar:1999xd,Hoang:2003ns,Hoang:2005dk,Pineda:2001ra}.},
and thereby the NRQCD factorization is verified in $\eta_{Q2}$ electromagnetic decay.
The terms of $\Delta^{(2)}_{\rm reg}$ and $\Delta^{(2)}_{\rm lbl}$ in (\ref{NNLO-separation}) represent the contributions
from the regular and `light-by-light' Feynman diagrams, which are illustrated in Fig.~\ref{feynman-fig-etac2}.

Furthermore, we can organize the $\Delta^{(2)}_{\rm reg}$ and $\Delta^{(2)}_{\rm lbl}$ according to the color structure,
\bqa
\label{delta-reg}
\Delta^{(2)}_{\rm reg}=C_F^2 s_A+C_F C_A s_{NA}+n_L C_F T_F s_L+n_H C_F T_F s_H,
\eqa
where
\bqa
\label{delta-reg-1}
s_A&=&-5.8455,\;\;\;
s_{NA}=-4.3701,\nn\\
s_L&=&1.4464,\;\;\;\;\;\; s_H=0.0161,
\eqa
and
\bqa
\label{delta-lbl}
\Delta^{(2)}_{\rm lbl}=(0.0002+0.0056\,i)n_H C_F T_F+(0.2136-0.0082\,i)C_F T_F \sum_{i}^{n_L}\frac{e_i^2}{e_Q^2}.
\eqa

By setting the renormalization scale $\mu_R=m_Q$ and factorization scale $\mu_F=1$ GeV, we get the radiative corrections to $\mathcal{C}_{1,1}$ at various perturbative orders,
\bqa
\label{sdc-per-order}
\mathcal{C}_{1,1}=\frac{4\sqrt{6}\pi}{3\sqrt{m_Q}}\alpha e_Q^2(1-0.62\alpha_s-r\alpha_s^2),
\eqa
where $r=2.12+0.0005\,i$ for $\eta_{c2}$ and $r=1.11+0.005\,i$ for $\eta_{b2}$. We find that
both the $\mathcal{O}(\alpha_s)$ and $\mathcal{O}(\alpha_s^2)$ corrections to the helicity amplitude
are negative as well as moderate. It seems that the perturbative expansion for $\eta_{Q2}\to 2\gamma$ exhibits a decent convergence,
however as will be found, the radiative corrections accurate up to $\mathcal{O}(\alpha_s^2)$ change the LO decay width considerably.

For completeness, it is necessary to deduce the explicit expression of decay width. Applying the formula Eq.~(\ref{decay-rate-1}) and the expression of helicity amplitude in Eq.~(\ref{helicity-amp-explicit}), we readily obtain the decay width of $\eta_{Q2}\to 2\gamma$ through $\mathcal{O}(\alpha_s^2)$
\bqa
\label{decay-rate-explicit}
\Gamma_{\gamma\gamma}(\eta_{Q2})&=&\frac{4\pi\alpha^2 e_Q^4}{15}\frac{|\langle0|\chi^{\dagger}{\mathcal K}_{^1D_{2}} \psi(\mu_F)|\eta_{c2}\rangle|^2}{m_Q^6}\bigg[1+\frac{\alpha_s}{\pi}2 C_F \Delta^{(1)}+\frac{\alpha_s^2}{\pi^2}
\bigg(C_F^2\Delta^{(1)2}+\nn\\
&&C_F\frac{\beta_0}{2}\Delta^{(1)}\ln\frac{\mu_R^2}{m_Q^2}
+2{\rm Re}\Delta^{(2)}
\bigg)
\bigg],
\eqa
where the symbol ${\rm Re}$ signifies the real part of the argument.

\section{phenomenology\label{sec-result-and-discussion}}
\subsection{predictions for the decay width}

To make concrete prediction, we first choose the input parameters.
We take the heavy quark mass to be $m_c=1.68$ GeV and $m_b=4.78$ GeV,
which correspond to the two-loop charm quark and bottom quark pole masses converted from the corresponding $\overline{\rm MS}$ masses~\cite{Chetyrkin:2000yt}.
We evaluate the electromagnetic coupling constant as $\alpha(2m_c)\approx\frac{1}{132}$ and $\alpha(2m_b)\approx\frac{1}{131}$ by the formulas in Ref.~\cite{Erler:1998sy}, and evaluate $\alpha_s$ at each energy
scale by {\tt RunDec}~\cite{Chetyrkin:2000yt}.

The NRQCD LDME is related to the second derivative of the 1D radial wave function at the origin in Eq.~(\ref{nrqcd-potential}),
which are well determined by the nonrelativistic potential model.
In this work, we take the $\mathcal{R}^{''}_{D}(0)$ from the Cornell potential model~\cite{Eichten:1995ch,Eichten:2019hbb}, and
determine the LDMEs as
\bqa
|\langle0|\chi^{\dagger}{\mathcal K}_{^1D_{2}} \psi|\eta_{c2}\rangle|^2\approx\frac{5N_c}{8\pi}\times0.0329=0.0196\;{\rm GeV}^7,\nn\\
|\langle0|\chi^{\dagger}{\mathcal K}_{^1D_{2}} \psi|\eta_{b2}\rangle|^2\approx\frac{5N_c}{8\pi}\times0.8394=0.5010\;{\rm GeV}^7.
\eqa

With these input parameters, we present our predictions for the decay widths of $\eta_{c2}/\eta_{b2}\to 2\gamma$ at various levels of accuracy in $\alpha_s$ in Tab.~\ref{table-numerical-prediction-cornell}. The uncertainties affiliated with the decay width are estimated by varying $\mu_R$ from $m_Q$ to $2m_Q$ with the central values evaluated at $\sqrt{2}m_Q$.~\footnote{In this work, we do not consider the uncertainty originating from the heavy quark pole mass, which is expected to be considerably greater than that from varying the renormalization as well as the NRQCD factorization scales. For more discussion about the heavy quark pole mass, we refer the
interested readers to Refs.~\cite{Marquard:2015qpa,Kataev:2015gvt,Ayala:2019hkn,Mateu:2017hlz}}
From the table, we have several observations. First, the NNLO decay width is much smaller than the LO one for the channel $\eta_{c2}\to 2\gamma$, which is accounted for by the sizeable and negative $\mathcal{O}(\alpha_s)$ and $\mathcal{O}(\alpha_s^2)$ radiative corrections. Second, the decay width is insensitive to the factorization scale $\mu_F$.
Third, the decay width of $\eta_{b2}\to 2\gamma$ is considerably smaller than the case of $\eta_{c2}$, which is mainly caused by the heavier quark mass and smaller electric charge for bottom quark.

We can also predict the branching ratio of $\eta_{Q2}\to 2\gamma$. According to current theoretical computation, $\eta_{c2}$ decay predominately
through the electric $E1$ transition and hadronic decay.
If assuming $\eta_{Q2}$ decay is saturated by these two decay patterns,
 we can approximate the total decay width through~\cite{Fan:2009cj}
\bqa
\Gamma_{\rm total}(\eta_{Q2})\approx \Gamma(\eta_{Q2}\to {\rm LH})+\Gamma(\eta_{Q2}\to \gamma h_Q),
\eqa
where LH denotes the abbreviation for light hadrons.
The hadronic decay width of $\eta_{Q2}$ up to NLO has been known for a long time~\cite{Fan:2009cj}, and the prediction for
electromagnetic $E1$ transition of $^1D_2\to^1P_1$ from Cornell potential model can be found in Refs.~\cite{Eichten:2002qv,Barnes:2005pb}.
Thus, we readily obtain the total decay width
$\Gamma_{\rm total}(\eta_{c2})=142.1+303.0=445.1$ keV and $\Gamma_{\rm total}(\eta_{b2})=3.8+25.3=29.1$ keV, where
we have reevaluated the hadronic decay width with the parameters selected in this work. Consequently, the branching ratio of $\eta_{Q2}\to 2\gamma$ is illustrated in Tab.~\ref{table-numerical-prediction-cornell}. It is remarkable that ${\rm Br}(\eta_{b2}\to 2\gamma)$ is much smaller than ${\rm Br}(\eta_{c}\to 2\gamma)$, which renders the search for $\eta_{b2}$ through its electromagnetic decay quite challenging.

\begin{table}
\caption{NRQCD predictions for the decay width of $\eta_{Q2}\to 2\gamma$ at various levels of accuracy in $\alpha_s$.
We take the two-loop quark pole masses $m_c=1.68$ GeV and $m_b=4.78$ GeV.
The NRQCD LDMEs are evaluated by the Cornell potential model.
The errors are estimated by sliding the renormalization scale $\mu_R$ from $m_Q$ to $2m_Q$ with center value $\mu_R=\sqrt{2}m_Q$.
By taking the total decay width as $\Gamma_{\rm total}(\eta_c)\approx 445.1$ keV and $\Gamma_{\rm total}(\eta_b)\approx 29.1$ keV,
we also present the branching ratio ${\rm Br}(\eta_{Q2}\to 2\gamma)$. }
\begin{center}
\label{table-numerical-prediction-cornell}
\begin{tabular}{|c|c|c|c|c|c|c|c}
\hline
$\Gamma_{\gamma\gamma}$ & \multicolumn{2}{c}{} &\multicolumn{2}{|c}{$\mu_F=1$ GeV}   & \multicolumn{2}{|c|}{$\mu_F=m_Q$ }\\
\cline{2-5}\cline{6-7}
in unit of eV  & LO & NLO  & NNLO &{\rm Br}($\eta_{Q2}\to 2\gamma$)   & NNLO &{\rm Br}($\eta_{Q2}\to 2\gamma$)  \\
\hline
$\eta_{c2}\to 2\gamma$ & $8.28$ & $5.68^{+0.25}_{-0.34}$ & $2.61^{+0.70}_{-1.04}$  & $5.9^{-2.3}_{+1.6}\times 10^{-6}$ &$2.11^{+0.81}_{-1.23}$& $4.7^{-2.7}_{+1.8}\times 10^{-6}$ \\
\hline
$\eta_{b2}\to 2\gamma$  & $0.025$ & $0.019_{-0.001}^{+0.001}$ &  $0.014_{-0.001}^{+0.001}$ & $4.7^{-0.4}_{+0.3}\times10^{-7}$ &$0.011_{-0.002}^{+0.001}$& $3.9^{-0.5}_{+0.4}\times10^{-7}$\\
\hline
\end{tabular}
\end{center}
\end{table}

\subsection{$\eta_{Q2}$ production at colliders}

In the following, we will evaluate the production cross section of $\eta_{Q2}$ at B factory and LHC.
The $\eta_{c2}$ associated production with a photon at B factory has been studied in Ref.~\cite{Li:2009ki}, and the
corresponding cross section at lowest order in $\alpha_s$ expansion is given
\bqa
\sigma(e^+e^-\to\eta_{c2}+\gamma)=\frac{80\pi\alpha^3 e_c^4 (1-4m_c^2/s)}{s^2 m_c^5}|\mathcal{R}''_D(0)|^2,
\eqa
where $\sqrt{s}=10.58$ GeV is the center-of-mass (CM) energy at B factory. With the aforementioned input parameters,
we immediately arrive at $\sigma(e^+e^-\to\eta_{c2}+\gamma)=1.49$ fb. Due to the small cross section and branching ratio,
it seems impossible to detect $\eta_{c2}$ through its electromagnetic decay at B factory.

Now we turn to the hadron collider LHC, where amount number of
quarkonia can be produced due to the considerable production cross section, e.g., the cross section of $\eta_c$ can reach
around $0.5 \mu$b  through single parton scattering (SPS)~\cite{Zhang:2014ybe}. For $pp\to \eta_{Q2}+X$, the differential cross section through SPS can be factorized as
\bqa
\label{cross-section-LHC}
d\sigma(pp\to\eta_{Q2}+X)&=&\sum_{i,j}\int dx_1 dx_2 f_{i/p}(x_1)f_{j/p}(x_2)d\hat{\sigma}(i+j\to c\bar{c}(^1D_2)+X)\nn\\
&&\times\frac{|\langle0|\chi^{\dagger}{\mathcal K}_{^1D_{2}} \psi(\mu_F)|\eta_{Q2}\rangle|^2}{m_Q^7},
\eqa
where we have neglected the color-octet contribution, $f_{i/p}(x)$ represents the parton distribution function of a proton, and $d\hat{\sigma}$ denotes the partonic cross section.
Since the gluon distribution is overwhelming in the proton at small momentum fraction $x$, we expect that
the gluon scattering will denominate the cross section. Thus, it is reasonable for us to consider gluon-gluon partonic scattering to
estimate the cross section of $\eta_{Q2}$. In addition, we will carry out the partonic cross section $d\hat{\sigma}$ at lowest order in $\alpha_s$.
For concreteness,
we consider the production of $\eta_{Q2}$ at LHCb detector, where a kinematic constraint on the longitudinal rapidity of $\eta_{Q2}$ $4.5>y>2$ is implemented.
We further take a transverse momentum $P_T$ cut for the $\eta_{Q2}$ to guarantee the validity of the factorization formula (\ref{cross-section-LHC}).
In our computation, we employ {\it CTEQ14} PDF sets~\cite{Dulat:2015mca} for the proton PDF.

\begin{table}
\caption{The cross sections of $\eta_{c2}$ and $\eta_{b2}$ at the lowest order in $\alpha_s$ expansion at LHCb. By taking the CM colliding energy $\sqrt{s}=7$ TeV and $13$ TeV, we evaluate the integrated cross sections with $\eta_{Q2}$ longitudinal rapidity constraint $4.5>y>2$ and transverse momentum $P_T$ cut. }
\begin{center}
\label{table-pp-etaQ2-cornell}
\begin{tabular}{|c|c|c|c|c|c|c|c|}
\hline
 & \multicolumn{3}{c|}{$\sqrt{s}=7$ TeV} & \multicolumn{3}{c|}{$\sqrt{s}=13$ TeV}\\
\hline
\diagbox{\quad\quad$\sigma$\\ in unit of nb}{$P_T$ cut\;\;\;\;\;\;\;\;} & $P_T>2m_Q$ & $P_T>3m_Q$ & $P_T>4 m_Q$ &  $P_T>2m_Q$ & $P_T>2m_Q$ & $P_T>4 m_Q$ \\
\hline
 $\sigma(pp\to \eta_{c2}+X)$ & 24.7 & 5.1 & 1.4 & 47.8 & 10.5 & 3.0 \\
 \hline
 $\sigma(pp\to \eta_{b2}+X)\times 10^3$ & 8.7 & 1.5 & 0.3 & 21.5 & 4.0 & 1.0 \\
 \hline
\end{tabular}
\end{center}
\end{table}
In Tab~\ref{table-pp-etaQ2-cornell}, we present the theoretical predictions for the cross section of $\eta_{Q2}$ at two benchmark CM energy $\sqrt{s}=7$ TeV and $13$ TeV  with various transverse momentum cutoffs for $\eta_{Q2}$.
From the table, we find the cross section of $\eta_{b2}$ is smaller than that of $\eta_{c2}$ by roughly three order of magnitude.
Combing the luminosity at LHCb, we can estimate the number of events for $\eta_{Q2}$ production. With the integrated luminosity
$\mathcal{L}=10\, {\rm fb}^{-1}$ at each CM energy, there are $10^7-10^8$ $\eta_{c2}$ and $10^4-10^5$ $\eta_{b2}$ event produced at LHCb. Therefore, LHCb will be an ideal platform to probe $\eta_{Q2}$.
Furthermore, multiplying the branching ratio of $\eta_{Q2}\to 2\gamma$, we predict that there are several hundreds of
double-photon events through $pp\to \eta_{c2}\to 2\gamma$, which is a promising channel to probe this undiscovered
charmonium. In contrast, the $\eta_{b2}$ is proved to be hard to detect through its electromagnetic decay at LHCb.

\section{Summary\label{sec-summary}}
Applying the NRQCD factorization formalism, we evaluate the $\eta_{Q2}$ electromagnetic decay into double photons up to $\mathcal{O}(\alpha_s^2)$ radiative corrections. For the first time, we scrutinize the validity of the NRQCD factorization for D-wave quakonium decay at NNLO. Both the $\mathcal{O}(\alpha_s)$ and $\mathcal{O}(\alpha_s^2)$ corrections to the decay width of $\eta_{Q2}\to 2\gamma$ are negative.  Although the radiative corrections to the helicity amplitude are moderate, the
corrections change the LO decay width significantly, especially for $\eta_{c2}$. By assuming $\eta_{Q2}$ decay is saturated by  the 
electric $E$1 transition and the hadronic decay, we obtain the branching ratios ${\rm Br}(\eta_{c2}\to 2\gamma)\approx 5\times 10^{-6}$ and
${\rm Br}(\eta_{b2}\to 2\gamma)\approx 4\times 10^{-7}$. We have also studied the $\eta_{Q2}$ production at LHCb. By imposing kinematic
restriction on the longitudinal rapidity and transverse momentum of $\eta_{Q2}$, we predict the cross sections to be $2-50$ nb  for $\eta_{c2}$ and
$1-22$ pb for $\eta_{b2}$ for various transverse momentum cutoffs. Thus, it is promising to observe $\eta_{c2}$ through its electromagnetic decay at LHCb, while quite challenging to detect $\eta_{b2}$ at current integrated luminosity.



\end{document}